\documentclass[aps,prd,groupedaddress,superscriptaddress,showpacs]{revtex4}

\usepackage{epsfig}
\def\S12{\mathrm{S}_{1,2}}

\newcounter{allequation}

\begin{document}


\title{Bounds on the mass of the \lowercase{$b'$} quark, revisited}

\author{ S. M. Oliveira}
\email[]{smo@cii.fc.ul.pt}
\affiliation{Centro de F\'\i sica Te\'orica e Computacional,
   Faculdade de Ci\^encias, Universidade de Lisboa,
   Av. Prof. Gama Pinto 2, 1649-003 Lisboa, Portugal}

\author{R. Santos}
\email[]{rsantos@cii.fc.ul.pt}
\affiliation{Centro de F\'\i sica Te\'orica e Computacional,
   Faculdade de Ci\^encias, Universidade de Lisboa,
   Av. Prof. Gama Pinto 2, 1649-003 Lisboa, Portugal}
\affiliation{ Instituto Superior de Transportes e
   Comunica\c{c}\~oes, Campus Universit\'ario
   R. D. Afonso Henriques, 2330-519 Entroncamento, Portugal}

\date{\today}

\begin{abstract}
Recent results from the DELPHI collaboration \cite{DELPHI_NOTE}
led us to review the present bounds
on the $b'$ quark mass. We use all available experimental data
for $m_{b'} > 96$ GeV to constrain the $b'$ quark mass
as a function of the Cabibbo-Kobayashi-Maskawa elements
in a sequential four generations model.
We find that there is still room for a $b'$ with a mass
larger than 96 GeV.
\end{abstract}


\maketitle

\section{Introduction}

It has long been known that a sequential fourth generation
within the Standard Model (SM) needs both quarks and leptons.
Half a generation would
imply that the gauge anomalies associated with triangle
diagrams would not cancel. It is also known 
\cite{Hagiwara:fs} that
SLC, and then LEP have set a bound
on the number of light neutrinos ($ m_{\nu} \, < \, M_Z/2$),
which is indisputably equal to three. This bound applies to
all new fermions that couple to the Z and one has to be
extremely open minded to accept a fourth neutrino
with a mass larger than around 45 GeV.  
Thus, there seems to be
no strong motivation for the search of a sequential
fourth generation (for a review see \cite{Frampton:1999xi}).
So why look for it?

Despite the strength of the previous arguments one should try
to experimentally exclude the existence of a fourth generation.
In fact such evidence does not yet exist.
The most recent precision electroweak results
\cite{Novikov:2001md}
allow a sequential
fourth generation if the quark masses
are not too further apart$^1$.  
\footnotetext[1]{This result is a strong bound on
the mass difference of a possible fourth generation.
Nevertheless, it should be noticed that the authors
assume no mixing of the extra families with the SM ones.}
The same results also disfavour
a degenerate fourth family if both the leptonic and hadronic
sector are degenerate. This is in agreement
with the conclusions of Erler and Langacker
\cite{Hagiwara:fs}. However, as discussed
in ref. \cite{Frampton:1999xi}, there are several
reasons to keep investigating this subject
starting with the fact that precision
results vary with time.
In ref. \cite{Frampton:1999xi} it can be seen 
that even if one takes a degenerate
fourth family of quarks with 150 GeV masses, it is enough
to choose a non-degenerate family of leptons
with masses of 100 GeV and 200 GeV and a Higgs
mass of 180 GeV for the discrepancy with experimental data
to fall from roughly
three to two standard deviations$^2$.  
\footnotetext[2]{Notice that we make no assumptions
on the values of the masses and couplings of the leptonic
sector of the model.} Moreover, it is clear
that any new physics
will also influence these results.

It was shown in refs. \cite{Frampton:1999xi,Arhrib:2000ct}
that the mass range $|m_{t'}-m_{b'}| \leq 60 \, \text{GeV}$, where
$t'$ and $b'$ are the fourth generation quarks,
is consistent with all available precision electroweak data.
This range enable us to say that even if 
$m_{b'}>m_{t'}$, the decay $b' \rightarrow t' \, W$
is forbidden. The decay $b' \rightarrow t' \, W^*$
although allowed, is phase space suppressed \cite{Sher:1999ae}
and
consequently extremely small in the mass range under study
(from now on we consider $m_{b'} < m_{t'}$).
Experimental data allow us 
to go only up to $m_{b'}$ close to $180$ GeV. Hence,
the $b'$ can not decay to a top quark.
Furthermore,
while some recent studies \cite{Yanir:2002cq,Huang:2000xe}
have constrained the Cabibbo-Kobayashi-Maskawa (CKM) elements
of the fourth generation, they
do not influence our results. Nevertheless we will take
into account the $2 \sigma$ bound 
$|V_{tb}|^2 + 0.75 |V_{t'b}|^2 \leq 1.14$  \cite{Yanir:2002cq}
coming
from $Z \rightarrow b \bar{b}$ to constrain the CKM 
element $V_{cb'}$ as a function of the $b'$ mass.

Present experimental bounds on the $b'$ mass
above 96 GeV
suffer from the drawback of assuming a
100 \% branching ratio for a specific
decay channel. As stated before the strongest
bound on the $b'$ mass comes from LEP \cite{Decamp:1989fk} and 
is $m_{b'}\, > \, 46$ GeV. Here all $b'$ decays were considered.
There are presently three bounds on the $b'$ mass
for $m_{b'}\, > \, 96$ GeV.
The first one \cite{Affolder:1999bs}, $m_{b'}\, > \, 199$ GeV, assumes
that $Br(b' \rightarrow b \, Z)=100 \%$. 
We will drop this condition and use instead their plot of
$\sigma (p \, \bar{p} \rightarrow b' \bar{b'} + X) \times
Br^2(b' \rightarrow b \, Z)$ as a function of the $b'$ mass.
The second one \cite{Abachi:1995ms} $m_{b'}\, > \, 128$ GeV, is 
based on the data collected in the top quark search. Because
the D0 collaboration looked for $t \rightarrow b \, W$, the analysis
can be used to set a limit on 
$\sigma (p \, \bar{p} \rightarrow b' \bar{b'} + X) \times
Br^2(b' \rightarrow c \, W)$. By doing so we assume that the
$b$ and $c$ quark masses are negligible and that
$\sigma (p \, \bar{p} \rightarrow b' \bar{b'}) \approx
\sigma (p \, \bar{p} \rightarrow t \bar{t})$. The obtained
limit
$m_{b'}\, > \, 128$ GeV assumes
$Br(b' \rightarrow c \, W) = 100 \%$.
The third bound is from CDF \cite{Abe:1998ee} and is based on
the decay $b' \rightarrow b \, Z$ followed by the search for
$Z \rightarrow e^+ \, e^-$ with displaced
vertices. Their excluded region is inside a rectangle
in the lifetime $(c \, \tau)$, $m_{b'}$ plane with
$\, 9 \times 10^{-3} \enskip \text{cm}\, < \, c \, \tau \, < \, 12
\enskip \text{cm}$  
and $m_{b}+ M_Z \, < \, m_{b'}\, < \, 148$ GeV sides.
Hence, the excluded region depends heavily on the $b'$
lifetime. But, contrary to the top quark
which has a lifetime of around $10^{-24} \, s$, 
the lifetime of a sequential $b'$ quark is
expected to be extremely large, especially knowing
that we are considering  a heavy $b'$.
In fact, depending on the CKM values and on 
the $b'$ and $t'$ masses,
the decay length can be as large as $10^{-4}$ cm
or even $10^{-3}$ cm in extreme  cases.
Nevertheless,
in this model, it is very hard
to go beyond that value. It is worth mentioning that
even with this huge lifetime,
the $b'$ always decays inside the detector and
hadronization occurs before it decays.
Thus, the limit obtained
in \cite{Abe:1998ee} which on top of
what was said assumes $Br(b' \rightarrow b \, Z)=100 \%$
can not be used in our analysis.

Hence, we think it is worthwhile to reexamine the
limits on the $b'$ mass. We will use
the CDF and the D0 data which, together with
the new DELPHI data, is all that there is  available
for $m_{b'} > 96$ GeV.
We will draw exclusion plots in the plane
($R_{CKM}$, $m_{b'}$), where
$R_{CKM}=|\frac{V_{cb'}}{V_{tb'} \, V_{tb}}|$,
from 96 GeV
to 180 GeV without assuming a definite value for the
branching ratios of specific channels.
In some regions it is possible to combine all experimental
data allowing a larger
exclusion area. Notice that the use of the $R_{CKM}$ 
variable provides a new way to look at the experimental results.
This variable
enable us to actually use and combine all the available
data. Moreover, the new form in which the results are 
presented
will serve as a guide to future experiments since
it is possible to know how far one has to go
to exclude the regions that are still allowed.

To end this section
we note that there is, at present, no bound on a sequential 
$2/3$ charged quark in the PDG but if we assume a 100 \% decay 
to cW the bound is again 128 GeV \cite{Abachi:1995ms}. 

The paper is organised as follows. In section II we define
the model and discuss the production and the decays of $b'$
quarks. In section III we combine the theoretical
and the experimental results to produce
exclusion plots in the parameter space.
Section IV summarises our results and conclusions.

\section{\lowercase{b'} production and decay}

There are several ways of extending the SM to accommodate
a fourth family of quarks and/or leptons. A review
of the different models in the literature
is available in \cite{Frampton:1999xi}. Obviously, 
the most natural and straightforward way to introduce a fourth
family in the SM is just to add a $(t', \, b')$ family with
the same quantum numbers and similar couplings
to all other known quarks. The same
can be done for the lepton sector$^3$.  
\footnotetext[3]{Now that it is finally
accepted that neutrinos have mass, the SM has
to be changed to accommodate this new feature. We do not
restrict ourselves to any specific mechanism
that generates the very high neutrino mass needed in SM4.}
This is called a sequential fourth generation
model and is sometimes referred to as SM4.
The resulting CKM matrix has a very similar
structure to the SM one. It is a 
$4 \times 4$
unitary matrix and it is assumed to be approximately
symmetric.
Besides the four new masses, there
are 9 additional parameters compared to the SM: 6 mixing
angles instead of 3 and 3 complex phases
instead of 1.
Because we are not
concerned with CP-violation
we take all CKM values to be real.
In the SM4, the CKM elements that are not
determined experimentally have more freedom 
due to the extra parameters introduced.
This model has been the subject of wide
study in the literature. Production cross
sections for lepton and hadron colliders
and $b'$ branching fractions were calculated long
ago.

At LEP, a pair of heavy quarks is produced
through the reaction $e^+ \, e^- \rightarrow  q \bar{q}$.
For consistency with the experimental analysis,
the process $e^+ \, e^-  \rightarrow b' \, \bar{b'}$
was calculated using \textsc{Pythia} \cite{Sjostrand:2000wi}, 
with initial state radiation (ISR), final state radiation (FSR)
and QCD corrections turned on.
We have cross checked the results using a simple
program with the formulas of refs. \cite{Jersak:sp} and  
\cite{Bernreuther:1991hy},
which also include QCD corrections and ISR.
Since the larger contribution to
the cross section comes from ISR we have double
checked by making use of the formulas presented
in \cite{Kuraev:hb}. The results agree
very well with the \textsc{Pythia} results. 
It should be noticed that near the threshold bound
states would surely be formed. Without a detailed
analysis of such bound states it is impossible
to evaluate whether their  contribution to the
cross section would be relevant or not. So, if
bound states do exist above the threshold, we are
assuming that they give a negligible contribution
to the cross section. Far away from the threshold
the problem ceases to exist and the results we will
show for hadron colliders are not affected by
this approximation.

The equivalent production reaction at the Tevatron 
is $p \, \bar{p} \rightarrow b' \, \bar{b'} + X$,
with the relevant processes being
$gg \, (q \, \bar{q}) \rightarrow b' \, \bar{b'}$.
Even though this cross section can not be found
in the literature it is generally
recognised that all massive quark pair production
cross section are very similar due to
its hadronic nature. The same is true for the
subsequent decays into leptons and for the detector efficiency.
Thus we can use the exact order $\alpha_s^3$ corrected cross
section for the production
of top quarks \cite{Laenen:1993xr}.
This approximation is used both by the CDF
and the D0 collaborations in their studies
on $b'$ production and decay. In \cite{Abachi:1995ms} 
it is also assumed that the final states are exactly
the same as the top quark ones. 
Notice that the error in calculating the hadronic cross
section is much larger than the corresponding error in
the leptonic one.
For $m_q = 100$ GeV the error is about 38 \%
falling to 12 \% for $m_q = 200$ GeV.  
This will be reflected in the exclusion plots.

All $b'$ decays were exhaustively studied
by Hou and Stuart 
in \cite{Hou:1988yu,Hou:1988sx,Hou:1989ty,Hou:1990wz}
and by Haeri, Eilam and Soni \cite{hes}.
Hou and Stuart have shown that the $b'$ is 
peculiar in the sense that 1-loop flavour
change decays (FCNC) can dominate over charged
current decays (CC). Depending on the values
of the  CKM
matrix elements and as long as the Higgs channel
remains closed, there are mainly two processes
in competition: $b' \rightarrow b \, Z$ and 
$b' \rightarrow c \, W$. As soon as the Higgs channel
opens the decay $b' \rightarrow b \, H$ can be as large
as $b' \rightarrow b \, Z$. Other decays like 
$b' \rightarrow b \, g$ and $b' \rightarrow b \, \gamma$
and three body decays give smaller
contributions but can sometimes be relevant.

The three body decays
$b' \rightarrow b \, e^+ \, e^-$,
$b' \rightarrow b \, \nu \, \bar{\nu}$ and 
$b' \rightarrow b \, q \, \bar{q}$, including box diagrams
were calculated in \cite{Hou:1989ty}. At that time, the top mass
was still unknown and the $t'$ was taken
to be much larger than the top mass. Under these conditions
and for the range of the $b'$ mass in study,
the sum of all three body decays could be as large as
$b' \rightarrow b \, g$. It could be even larger
for a ``small'' $t$ mass and a very large
$t'$ mass \cite{Hou:1989ty}. But
it turned out that the top mass is $\approx$ 175 GeV and 
electroweak precision
measurements force $m_{t'}$ to be close to $m_{b'}$
for the range of $b'$ mass under
consideration. In our case we estimate
all three body decays plus the decay $b' \rightarrow b \, \gamma$
to be smaller than  $b' \rightarrow b \, g$. Nevertheless,
because we want to make a conservative estimate we
will
take it to be as large as $b' \rightarrow b \, g$.

Using the unitarity of the CKM matrix,
its approximate symmetry $V_{t'b'} \, V_{t'b} \approx
V_{tb} \, V_{tb'}$, and taking  
$V_{ub'} \, V_{ub} \approx \, 0$
and $V_{cb} \approx \, 10^{-2}$
we can write all branching fractions
as a function of three quantities alone:
$R_{CKM}$, $m_{t'}$ and
$m_{b'}$.
Notice that the two last conditions 
do not play a significant role in the final result.
Using a very
large value like for instance
$V_{ub'} \, V_{ub} \approx \, 10^{-4}$ gives
a contribution much less than 1 \% to
the $b' \rightarrow b \, Z$ decay width.
The same is true when we relax the condition 
$V_{t'b'} \, V_{t'b} \approx
V_{tb} \, V_{tb'}$ near to a GIM cancellation
region. Relaxing this condition leads to
an increase by several orders of magnitude of the values
of the NC decay widths but they are always much smaller than
the CC decays in that region.

One-loop calculations of the NC $b'$ decays
were performed using the FeynArts and FeynCalc \cite{feyn}
packages for 
generating and computing the complete set of diagrams
and the LoopTools/FF \cite{ff} packages for the numerical analysis.
We have carried out several checks in the four generations
model following
\cite{Arhrib:2000ct,Hou:1988yu,Hou:1988sx,Hou:1989ty,Hou:1990wz}
and in the SM against \cite{Mele,Eilam}. 
We have found full agreement in both cases.

The branching ratios depend on three quantities alone
and 96 GeV $\lesssim m_{b'}\lesssim$ 180 GeV. So, we just have
to decide on what
values of $R_{CKM}$ and $m_{t'}$ to use. Since we know that
$m_{t'}$ is limited by precision data we will study two extreme
cases $m_{t'} \, = \, m_{b'}+ 50$ GeV and the almost degenerate case  
$m_{t'} \, = \, m_{b'}+ 1$ GeV. In the exclusion plots
$R_{CKM}$ is a free parameter and so no assumptions on its
variation range were made.
However, there is a hint on its most significant values
coming from the fact that
the competing NC and CC
cross at $ 10^{-3} \lesssim R_{CKM} \lesssim 10^{-2}$.
We will come back to this point later.

\begin{figure}[htbp]
  \begin{center}
    \epsfig{file=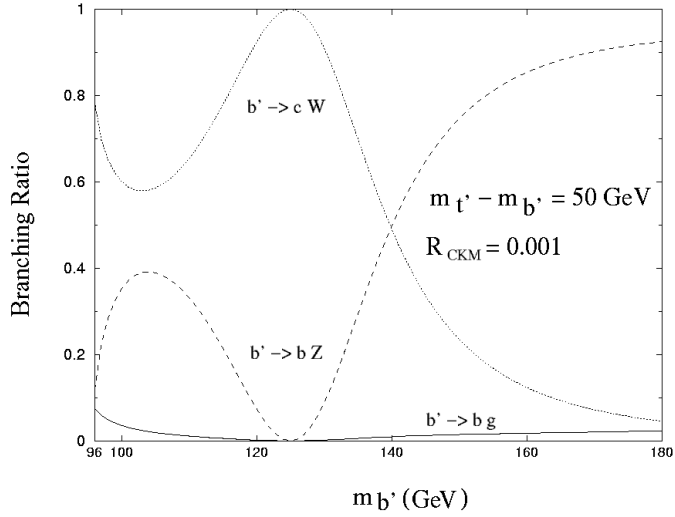,width=12 cm,angle=0}
    \caption{Branching ratios as a function of the $b'$ mass.
            The Higgs channel is closed.
            $R_{CKM}=0.001$ and
            $m_{t'} \, = \, m_{b'}+ 50 \, \text{GeV}$.
            The dashed line is
            $b'\rightarrow \, b \, Z$; the full line is
            $b'\rightarrow \, b \, g$ and the dotted line is
            $b'\rightarrow \, c \, W$.}
    \label{fig_new1}
  \end{center}
\end{figure}

In fig. \ref{fig_new1} we present the branching
ratios as a function of the $b'$ mass with $R_{CKM}=0.001$
and  $m_{t'} - m_{b'}= 50$ GeV. The closer
to $m_{b'}=96$ GeV we are the larger
$b'\rightarrow \, b \, g$ gets due to phase space suppression
of the competing NC  $b'\rightarrow \, b \, Z$.
In fact, for an almost degenerate fourth family and 
small values of $R_{CKM}$, $b'\rightarrow \, b \, g$ can be the dominant
NC for  $m_{b'}=96$ GeV. As soon as one moves away from this value,
$b'\rightarrow \, b \, Z$ becomes the dominant NC. If the Higgs
channel is closed , for $m_{b'} \geq 97$ GeV, the competition
is always between $b'\rightarrow \, c \, W$
and $b'\rightarrow \, b \, Z$. As $m_{b'}$ rises so
does the NC except if the GIM mechanism gets in the way.
It can be clearly seen in the figure the GIM mechanism
acting for
$m_{b'} \approx 125$ GeV, that is, $m_{t'}-m_{t}=0$.
Then the NC rises again and the CC falls crossing at 140 GeV.
When $R_{CKM}$ grows so does  $b'\rightarrow \, c \, W$ and the
crossing point is shifted to the left. As the mass difference
tends to zero the GIM effect is shifted to  $m_{b'} \approx m_{t}$.

In fig. 2 we show the branching ratios as a function of 
$R_{CKM}$ with  $m_{b'} = 110$ GeV and 
$m_{t'} - m_{b'}=  1$ GeV. As
we already knew, the NC are favoured by small values
of $R_{CKM}$ because $R_{CKM}$ is a direct measure of the charged
currents. Again, when $m_{b'}$ grows so does $b'\rightarrow \, b \, Z$
and the crossing point is shifted to the left. The same happens
when $m_{t'} - m_{b'}$ decreases as explained above.

\begin{figure}[htbp]
  \begin{center}
    \epsfig{file=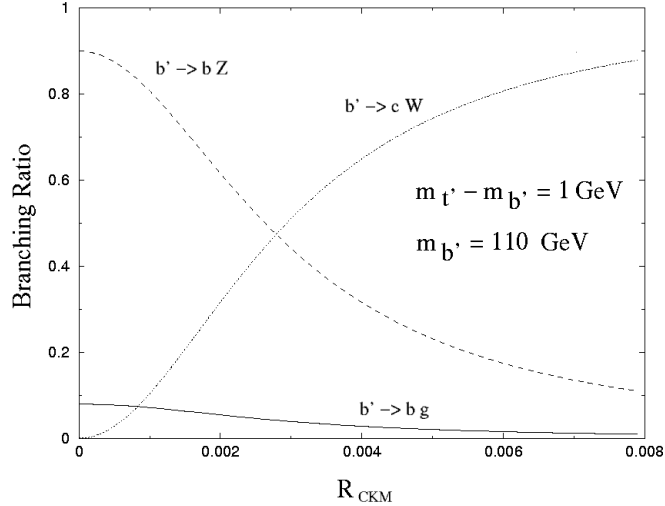,width=12 cm,angle=0}
    \caption{Branching ratios as a function of the $R_{CKM}$
            with $m_{b'} = 110$ GeV and
            $m_{t'} \, = \, m_{b'}+ 1$ GeV.
            The dashed line is
            $b'\rightarrow \, b \, Z$; the full line is
            $b'\rightarrow \, b \, g$ and the dotted line is
            $b'\rightarrow \, c \, W$. Higgs channel is closed.}
    \label{fig_new2}
  \end{center}
\end{figure}

\section{Results and discussion}

We are now in a position to draw exclusion plots
on the plane $(R_{CKM}, m_{b'})$ with $m_{t'}$
as a parameter.
Using the latest experimental data from the 
DELPHI collaboration
and the data from the CDF and D0 collaborations
together with the theoretical values of the cross
sections and the branching ratios we have drawn
the exclusion plots shown in the figures below.
The upper regions are excluded by the limits
on $Br_{b'\rightarrow \, c \, W}$ and the lower regions
by the limits on $Br_{b'\rightarrow \, b \, Z}$. 

\begin{figure}[htbp]
  \begin{center}
    \epsfig{file=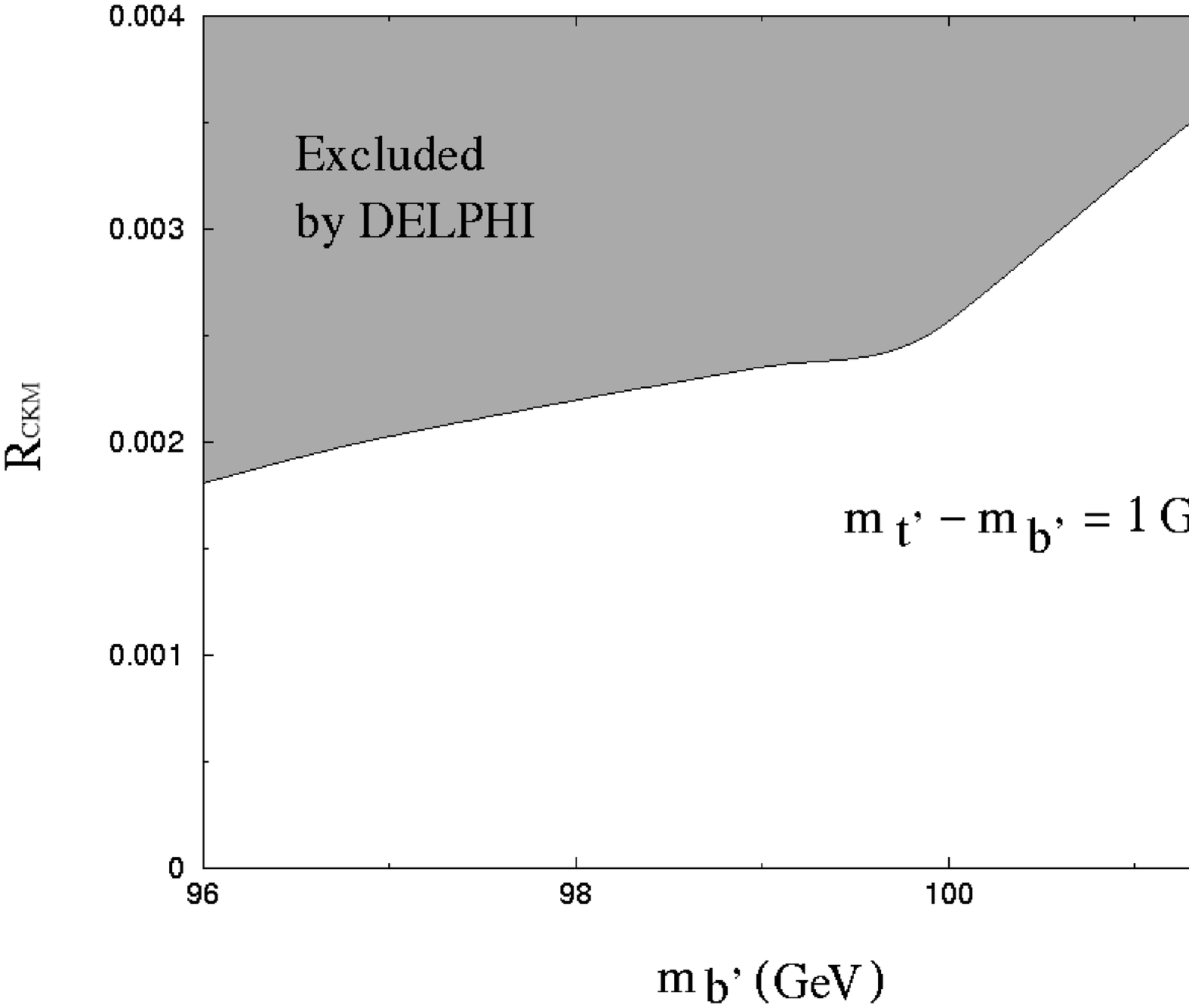,width=12 cm,angle=0}
    \caption{95 \% CL excluded region in the plane
            ($R_{CKM}$, $m_{b'}$) with
            $m_{t'}-m_{b'} = 1 \, \text{GeV}$,
            obtained from limits on
            $Br_{b'\rightarrow \, b \, Z}$
            and 
            $Br_{b'\rightarrow \, c \, W}$ (top).}
    \label{fig_lip1}
  \end{center}
\end{figure}

The results based on the DELPHI data,
are shown in figs. 3 and 4. 
The only difference between the two plots
is in the value of $m_{t'}$. 
It can be seen that as $m_{t'}-m_{b'}$
grows, the allowed region gets smaller.
This is because  $Br_{b' \rightarrow b \, Z}$
decreases with $m_{t'}$ due to  a GIM suppression
as long as $m_{t'}$ is smaller
than $m_{t}$ and $(m_{t'}-m_{t}) \rightarrow 0$. On
the contrary, $Br_{b' \rightarrow c \, W}$ does not depend
on the $t'$ mass.
Hence, as $m_{t'}$ grows, $Br_{b' \rightarrow c \, W}$
becomes dominant and the upper excluded region increases.

The reason why there isn't a lower bound 
close to 96 GeV in both figures
is because of the competing neutral currents. Close to
the $Z \, b$ threshold ($\approx$ 96 GeV),
$b' \rightarrow b \, g$ dominates over $b' \rightarrow b \, Z$
and the experimental bound on $Br_{b' \rightarrow b \, Z}$
becomes useless. As one moves away from the $Z \, b$ threshold,
$b' \rightarrow b \, Z$ becomes the dominant
neutral current. $Br_{b' \rightarrow b \, Z}$
falls less sharply with $m_{t'}$ than the other
neutral currents and that
explains why there is a lower bound for e.g. at
$m_{b'} = 100$ GeV in fig. 4 but not in fig. 3. 
After 102 GeV almost all values are allowed because
the experiments are not sensitive to those mass values. 
\begin{figure}[htbp]
  \begin{center}
    \epsfig{file=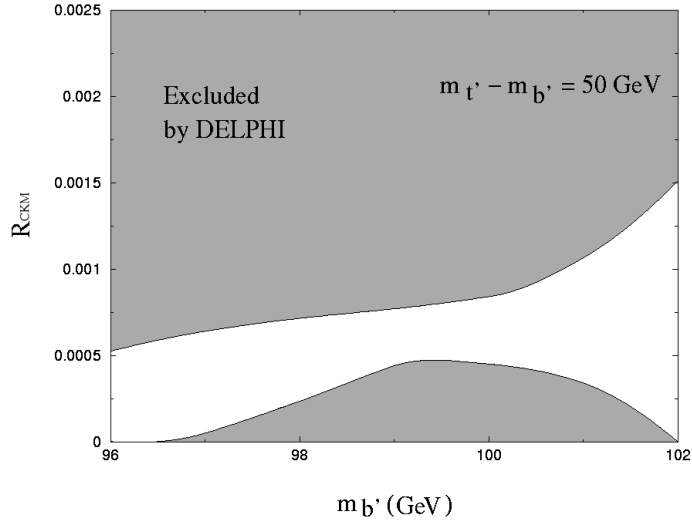,width=12 cm,angle=0}
     \caption{95 \% CL excluded region in the plane
            ($R_{CKM}$, $m_{b'}$) with
            $m_{t'}-m_{b'} = 50 \, \text{GeV}$,
            obtained from limits on
            $Br_{b'\rightarrow \, b \, Z}$ (bottom)
            and 
            $Br_{b'\rightarrow \, c \, W}$ (top).}
    \label{fig_lip2}
  \end{center}
\end{figure}

\begin{figure}[htbp]
  \begin{center}
    \epsfig{file=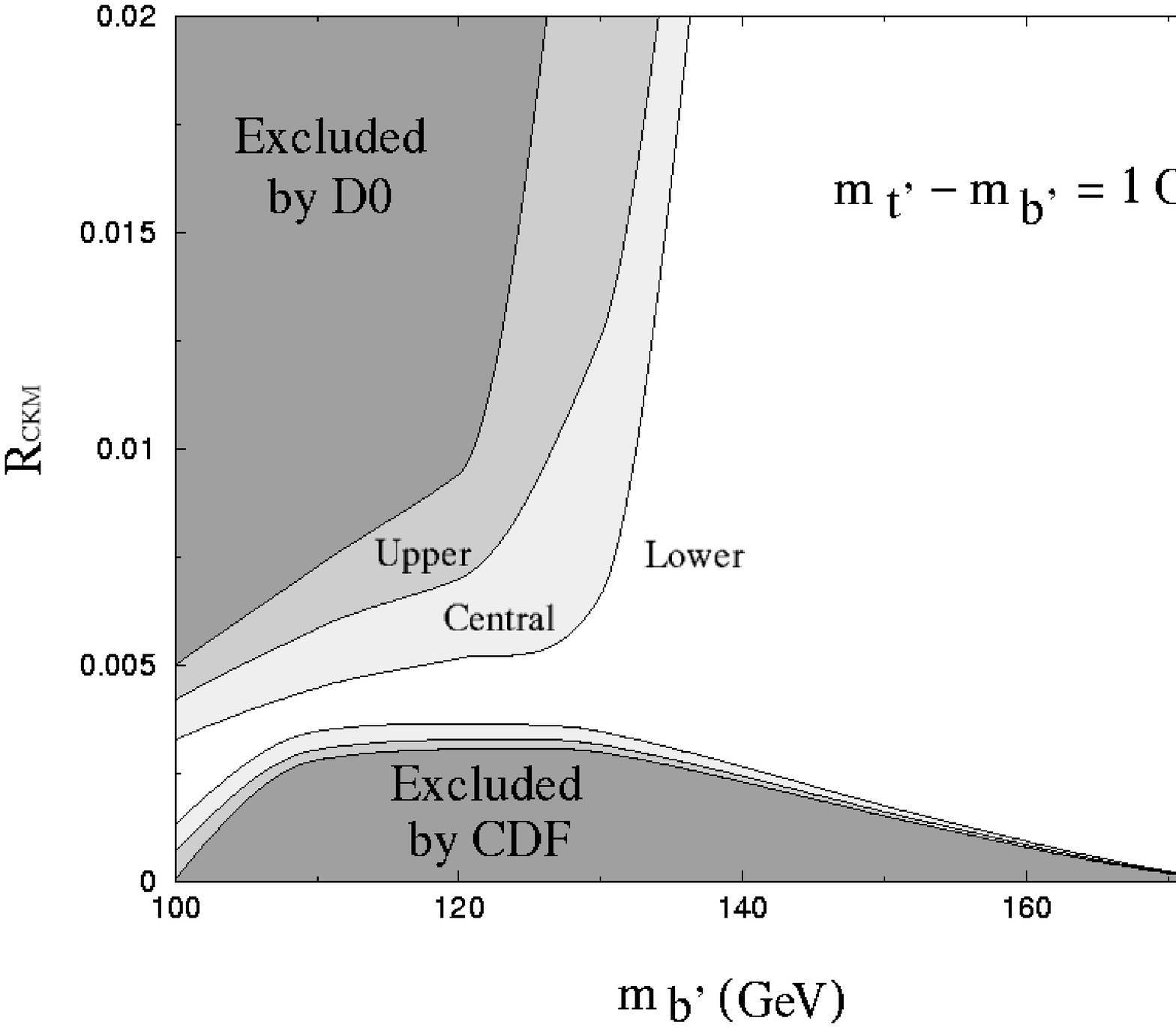,width=12 cm,angle=0}
     \caption{95 \% CL excluded region in the plane
            ($R_{CKM}$, $m_{b'}$) with
            $m_{t'}-m_{b'} = 1 \, \text{GeV}$,
            obtained from limits on
            $Br_{b'\rightarrow \, b \, Z}$ by the CDF coll. (bottom)
            and 
            $Br_{b'\rightarrow \, c \, W}$ by the D0 coll. (top).
            Upper, Central and Lower curves correspond to the values
            used for the $b'$ production cross-section.}
    \label{fig_cdf1}
  \end{center}
\end{figure}

\begin{figure}[htbp]
  \begin{center}
    \epsfig{file=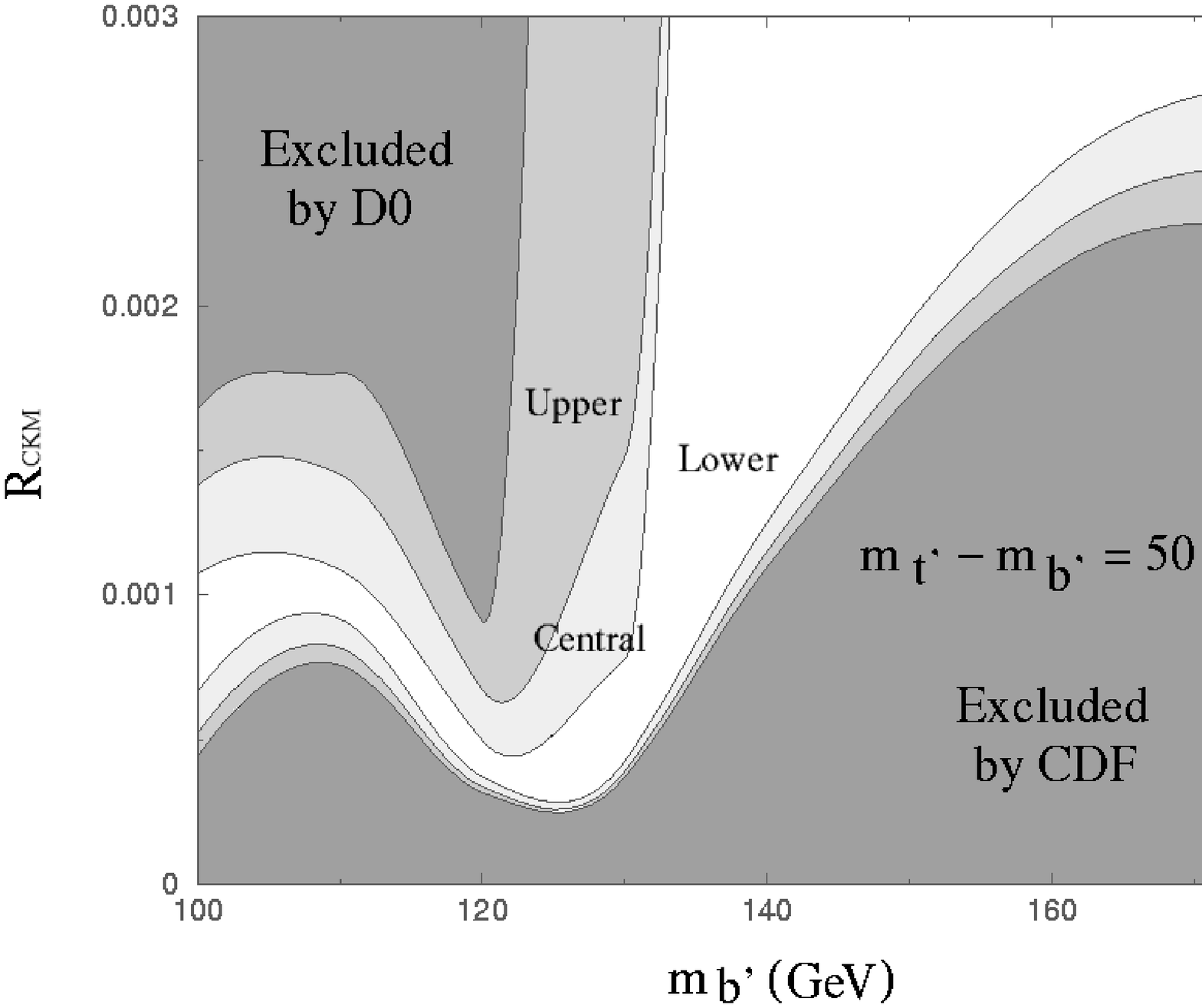,width=12 cm,angle=0}
     \caption{95 \% CL excluded region in the plane
            ($R_{CKM}$, $m_{b'}$) with
            $m_{t'}-m_{b'} = 50 \, \text{GeV}$,
            obtained from limits on
            $Br_{b'\rightarrow \, b \, Z}$ by the CDF coll. (bottom)
            and 
            $Br_{b'\rightarrow \, c \, W}$ by the D0 coll. (top).
            Upper, Central and Lower curves correspond to the values
            used for the $b'$ production cross-section.}
    \label{fig_cdf50}
  \end{center}
\end{figure}

In figs. 5 and 6 we show similar plots but using the CDF
and the D0 data. The D0 data is responsible for
excluding the upper regions because
it deals with CC as the CDF excludes the lower regions
due to the bounds on NC. The three curves marked upper,
central and lower are related with the theoretical
error bars in the $b'$ production cross section.
Again and for the same reason the excluded region grows
with $m_{t'}-m_{b'}$. This means that like the
constraints from precision electroweak data, the
experimental data also disfavours a fourth family
with a large mass difference between the two quarks.

\begin{figure}[htbp]
  \begin{center}
    \epsfig{file=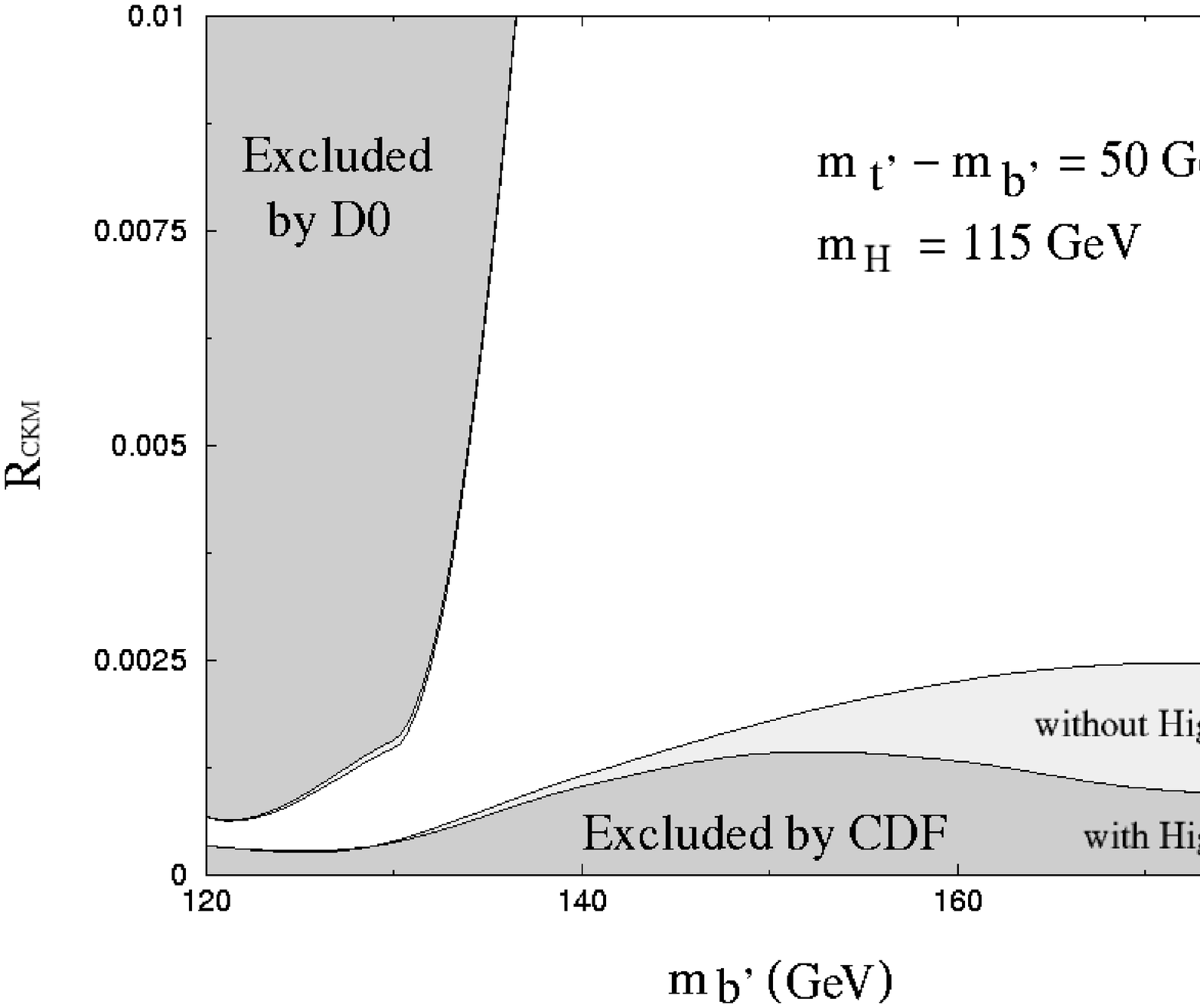,width=12 cm,angle=0}
     \caption{95 \% CL excluded region in the plane
            ($R_{CKM}$, $m_{b'}$) with
            $m_{t'}-m_{b'} = 50 \, \text{GeV}$,
            obtained from limits on
            $Br_{b'\rightarrow \, b \, Z}$ by the CDF coll.
            (bottom)
            and 
            $Br_{b'\rightarrow \, c \, W}$ by the D0 coll. 
            (top). The darker region is the excluded region
            with a Higgs boson of 115 GeV. Central values were
            taken for $b'$ production cross section.}
    \label{fig_cdfHiggs}
  \end{center}
\end{figure}

In some cases the allowed regions in the CDF/D0 and DELPHI
plots overlap and the excluded region grows. For instance,
considering $m_{b'} = 100$ GeV 
and $m_{t'}-m_{b'} = 50$ GeV we get for DELPHI
$4.5 \times 10^{-4} < R_{CKM} \, < \, 8.4 \times 10^{-4}$
and for CDF/D0 (lower)
$6.7 \times 10^{-4} < R_{CKM} \, < \, 1.1 \times 10^{-3}$.
Hence, the resulting excluded region is
$6.7 \times 10^{-4} < R_{CKM} \, < \, 8.4 \times 10^{-4}$.

With the bound 
$|V_{tb}|^2 + 0.75 |V_{t'b}|^2 \leq 1.14$  \cite{Yanir:2002cq}
and assuming $|V_{tb}| \approx 1$ it is possible
to limit the value of the matrix element 
$V_{cb'}$. For the same value of the $b'$ mass, $m_{b'} = 100$ GeV
we know
that $R_{CKM} \, < \, 8.4 \times 10^{-4}$ and so
\[
V_{cb'} \, < \, 8.4 \times 10^{-4} \sqrt{0.14/0.75} \,
\approx \,  3.6 \times 10^{-4} 
\]
with $m_{t'}= m_{b'} + 50  =  150 \, \text{GeV}$.
The bound gets weaker for smaller $m_{t'}$ \cite{Yanir:2002cq}.

Finally we show an exclusion plot with the Higgs channel
opened and a Higgs mass of 115 GeV. As we expected, the 
inclusion of the Higgs makes the excluded region
to shrink. By itself, the inclusion of one more
channel always diminishes the branching ratios and
consequently less values will be excluded.
Like $b' \rightarrow b \, Z$,  $b' \rightarrow b \, h$
is larger for small $R_{CKM}$ and large $m_{b'}$. Hence
in this region of parameter space it competes
with  $b' \rightarrow b \, Z$ and 
$b' \rightarrow c \, W$ making the allowed region
larger. For a detailed analysis of the so-called
cocktail solution see \cite{Arhrib:2000ct}.

\section{Conclusion}

In this work we have found the allowed $b'$ mass
as a function of the CKM elements of a four
generations sequential model. Using all available
experimental data for  $m_{b'}> 96$ GeV we have
shown that there is still plenty of room for a $b'$
with a mass larger than 96 GeV. We have also shown
that the allowed region decreases as  $m_{t'}$
increases. In fact, as the gap between the fourth generation quark masses
increases the allowed region shrinks. Notice that
this is in full agreement with the tendency of
a small mass gap, if not completely degenerated, 
favoured by the electroweak
precision measurements.

All plots show that $R_{CKM}$ is for sure
smaller than $\approx 10^{-2}$ and it can be as
small as  $\approx 10^{-4}$. This is not surprising
because this region is exactly where we expected
it to be. In fact, the CKM values we know so far
suggest that $V_{cb'} \approx 10^{-4}-10^{-3}$. If 
$V_{tb'} \approx 10^{-1}$ then a value of 
$R_{CKM}$ between  $10^{-2}$ and $10^{-4}$ is 
absolutely natural. 
Moreover, the limit we have obtained for
$V_{cb'}$ in the last section makes it
even more natural.

We know that the DELPHI analysis \cite{DELPHI_NOTE}
is being improved. In the near future we hope to reduce
very much the allowed region in figs. 3 and 4.
As far as we know there are no new results from the CDF
and the D0 collaborations improving their bounds.
For large $m_{t'}-m_{b'}$, and for 
some values of $m_{b'}$ the CDF/D0 limits
almost shrink the allowed region to zero. Hence,
a small improvement in the analysis could 
disallow a large region of the parameter space. 

As for the future, searches in hadron colliders
will have to wait for the RunII of the Tevatron
and for the Large Hadron Collider (LHC). The $b' \bar{b'}$
production cross section increases by roughly two orders
of magnitude at the LHC compared to the Tevatron. Thus
LHC will be a copious source of $b'$ pairs. With high
values for cross section and luminosity, if background is suppressed
exclusion plots can be drawn for a very wide range of $b'$
masses. However, we have to worry about two problems in
future searches. From the theoretical point of view we have
to take into account all the possible
hierarchies in mass, for instance one could
have $m_{t'} < m_{t} < m_{b'}$ or  $m_{t} < m_{t'} < m_{b'}$.
A careful study, including also the possibility of finding a
Higgs has to be done. From the experimental point of view
we have to know how the detectors will perform.

Nobody knows yet if there is going to be
a Next Linear Collider with
energies of $\sqrt{s}= 500$ GeV or  $\sqrt{s}= 1$ TeV.
NLC would allow us to go up $m_{b'}=250$ GeV or 
$m_{b'}=500$ GeV which is close to the perturbative
limit. Depending on the available
luminosity, and because a small background is expected,
we believe that the excluded region
would be very large, probably allowing the exclusion of
some values of  $m_{b'}$ regardless of
the values of the mixing angles.
However, if a Higgs boson is found
the excluded region will surely be smaller and will
depend on the mass and type of Higgs boson found.
For a detailed discussion on future searches see
\cite{Frampton:1999xi}. 

In summary we believe that there is still experimental
and theoretical work to be done to find or definitely
to exclude a sequential fourth generation of quarks
at the electroweak scale.

\begin{acknowledgments}
We thank A. Barroso and M. Pimenta for comments and suggestions
on the manuscript. We thank A. Onofre
for reading of the manuscript.
We thank our DELPHI/LIP collaborators and also M. Greco for discussions.

This work is supported
by Funda\c{c}\~ao para a Ci\^encia e Tecnologia under contract
POCTI/FNU/49523/2002. S.M.O. is supported by FCT
under contract SFRH/BD/6455/2001.

\end{acknowledgments}

\end{document}